\documentclass[iop,apjl]{emulateapj}

\newcommand{\igrjc}{IGR~J17480--2446\xspace}

\newcommand{\rxte}{\textit{RXTE}\xspace}
\newcommand{\integral}{\textit{INTEGRAL}\xspace}

\newcommand{\chandra}{\textit{Chandra}\xspace}
\newcommand{\xmm}{\textit{XMM-Newton}\xspace} 
\newcommand{\nustar}{\textrm{NuSTAR}\xspace} 

\newcommand{\np}[2]{\ensuremath{#1\,\times\,10^{#2}}}

\usepackage{graphicx} 
\usepackage{amsmath, amssymb, subfigure, ulem, xspace, wasysym} 

\usepackage{natbib} 
\usepackage[english]{babel} 

\bibpunct{(}{)}{;}{a}{}{,}

\shorttitle{SUBARCSECOND LOCATION OF \igrjc WITH \textit{ROSSI} XTE} 
\shortauthors{Riggio et al.}

\begin{document} 

\title{Subarcsecond location of \igrjc with \textit{Rossi} XTE}

\email{ariggio@oa-cagliari.inaf.it} 

\author{A. {Riggio}\altaffilmark{1,2}, L. {Burderi}\altaffilmark{2},
  T. {Di Salvo}\altaffilmark{3}, A. {Papitto}\altaffilmark{4},
  E. {Egron}\altaffilmark{2}, T. {Belloni}\altaffilmark{5},\\
  A. {D'A\`\i}\altaffilmark{3}, R. {Iaria}\altaffilmark{3},
  M. {Floris}\altaffilmark{6}, S. {Motta}\altaffilmark{5},
  V. {Testa}\altaffilmark{7}, M. T. {Menna}\altaffilmark{7},
  N. R. {Robba}\altaffilmark{3}}
 
\altaffiltext{1}{INAF-Osservatorio Astronomico di Cagliari, localit\`a
  Poggio dei Pini, Strada 54, 09012 Capoterra, Italy}

\altaffiltext{2}{Universit\`a~di Cagliari, Universit\`a~di Cagliari, SP
  Monserrato-Sestu km 0,7, 09042 Monserrato (CA), Italy}

\altaffiltext{3}{Dipartimento di Fisica, Universit\`a~di Palermo, Via
  Archirafi 36, Palermo, 90123 Italy}

\altaffiltext{4}{Institut de Ci\`encies de l’Espai (IEEC-CSIC), Campus
  UAB, Fac. de Ci\`encies, Torre C5, parell, 2a planta, 08193
  Barcelona, Spain}

\altaffiltext{5}{INAF-Osservatorio Astronomico di Brera, via
  E. Bianchi 46, 23807 Merate, Italy}

\altaffiltext{6}{CRS4-Center for Advanced Studies, Research and
  Development in Sardinia, Loc. Piscina Manna, Edificio 1, 09010 Pula
  (CA), Italy}

\altaffiltext{7}{INAF-Osservatorio Astronomico di Roma. Via Frascati,
  33, 00040 Monte Porzio Catone, Italy}

\date{} 

\begin{abstract} 
  On 2010 October 13, the X-ray astronomical satellite \textit{Rossi}
  XTE, during the observation of the newly discovered accretion
  powered X-ray pulsar \igrjc, detected a lunar occultation of the
  source.
  From knowledge of lunar topography and Earth, Moon, and spacecraft
  ephemeris at the epoch of the event, we determined the source
  position with an accuracy of 40 mas ($1\sigma$ c.l.), which is
  interesting, given the very poor imaging capabilities of \rxte
  ($\sim$ 1$^{\circ}$).
  For the first time, using a non-imaging X-ray observatory, the
  position of an X-ray source with a subarcsecond accuracy is
  derived, demonstrating the \textbf{neat} capabilities of a technique
  that can be fruitfully applied to current and future X-ray missions.
\end{abstract}

\keywords{Moon -- pulsars: general -- pulsars: individual
  (\objectname{\igrjc}) -- stars: neutron -- X-rays: binaries}

\section{Introduction} 
Since the dawn of the X-ray astronomy, the lunar occultation technique
has been used to study the position and structure of X-ray sources
\citep[see, e.g.,][and references therein]{Born_79}.
An X-ray mission like the ESA mission EXOSAT \citep{Taylor_EXOSAT_81}
was conceived to have, as part of the core program, the study of X-ray
sources using lunar occultations \citep{Born_79,Born_Debrunner_79}.
However, the development of X-ray mirrors with arcsecond resolution,
rendered, de facto, ineffective such possibility.
Indeed, the precision on the spacecraft ephemeris and the lunar
surface topography at that time, gave an uncertainty greater than the
imaging resolution of such optics.
A decade passed before \cite{Mereghetti_90} proposed again this
technique for missions like \xmm, but there is no trace in the last 20
years of literature of such an application.

\igrjc was detected for the first time on 2010 October 10 by the
IBIS-ISGRI instruments on board \integral \citep{Bordas_atel_10} in
the globular cluster Terzan 5.
While observed by the X-ray observatory \textit{Rossi} XTE (\rxte
hereafter), the source showed an eclipse, first attributed to the
companion star \citep{Strohmayer_atel_10a}.
A further investigation revealed that a very rare and serendipitous
event happened: the source was eclipsed by the Moon while observed by
\rxte \citep{Strohmayer_atel_10b}.

\cite{Pooley_atel_10}, on the basis of a \chandra observation of this
outburst, gave the most accurate source position, identifying \igrjc
with a quiescent LMXB previously detected in a \chandra observation of
Terzan 5 \citep{Heinke_06}.
\cite{Testa_atel_11} report the identification of a possible near-IR
counterpart through observations of the field with Adaptive Optics
systems before and after the discovery of the source. 
A detailed discussion will be presented in a forthcoming paper
(V. Testa et al. 2012, in preparation).
Very recently, \cite{Patruno_atel_12} report the identification of the
candidate counterpart of \igrjc in the optical bands in \textit{Hubble
  Space Telescope} 2003 archival observations whose position is in
agreement with the X-ray positions reported by \cite{Pooley_atel_10}
and this work.

In the following we will show how, taking advantage from this
fortunate event, it is possible to determine the position of \igrjc
with a subarcsecond precision.
The discussion is divided in two parts, where we explain the
occultation technique and provide a detailed analysis of the error
budget, respectively.

\section{The technique}\label{sec:technique} 
The precision of the lunar occultation technique depends on four
factors: (1) the precision on the position of the Moon, (2) the
precision on the spacecraft position, (3) the precision on ingress and
egress epochs, (4) and the precision of the lunar surface topography.

During the ingress and egress of a lunar eclipse, the occulted X-ray
source lays on the projection of the lunar rim on the plane of sky as
seen by the telescope (in this case \textit{Rossi} XTE).
The source position in the plane of sky is then given by the
intersection of these two profiles.
In general, the possible intersections are two, one of which can be
usually excluded, under some conditions, if a prior estimate of the
source position is available.

To reconstruct the profile of the lunar rim in the plane of sky, as
seen from \rxte at a given epoch, we first considered the position of
the spacecraft and of the Moon barycenter at a given epoch (in our
case ingress or egress epochs).
The space-time reference frame we adopted is a Cartesian coordinates
in a geocentric international celestial reference frame (ICRF/J2000.0)
for the position and terrestrial time (TT/TDT) for time.

\begin{figure} 
  \resizebox{\hsize}{!}{\includegraphics{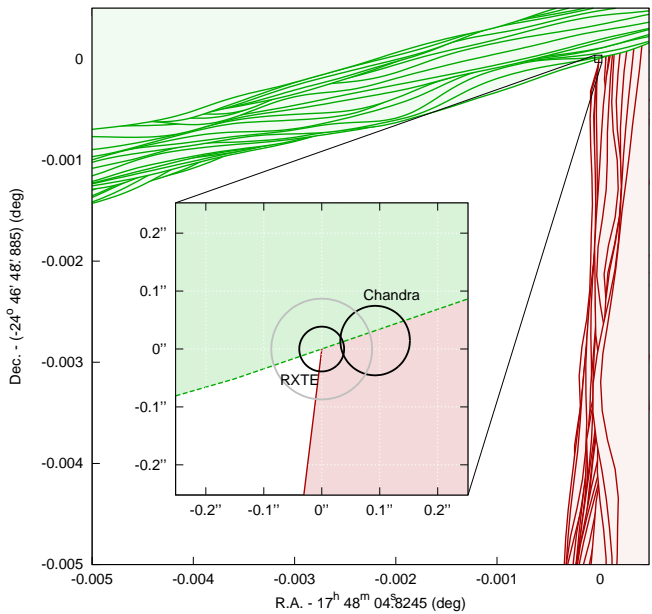}}
  \caption{Moon topography projection on the plane of sky as seen by
    \textit{Rossi} XTE at eclipse ingress (red curves) and egress (green
    curves).  Each curve is the projection on the plane of sky of a
    lunar profile at constant selenographic longitude. The envelope of
    these curves is the lunar rim as seen by \rxte.  In the inset the
    detail of the intersection between the two profiles determining
    our best source position is reported. Circles identify the
    \cite{Heinke_06} \chandra position and the position derived in
    this work (black $1\sigma$ c.l., gray $2\sigma$ c.l.),
    respectively.\label{fig:ecl_inset}}
\end{figure}

Moon position $\mathbf{r_{\leftmoon}}(t)$ and spacecraft position
$\mathbf{r_{XTE}}(t)$ are available as astrometric positions, i.e., as
a function of time in a reference frame centered on the Earth
barycenter.
Since the Moon astrometric and apparent positions (as seen from \rxte)
differ because of the travel time of X-ray photons to cover the
Moon--\rxte distance $d$, we must correct for this time, $\Delta t = d
/ c$, where $c$ is the speed of light.
The expression to evaluate the vector distance $\mathbf{d}$ is
reported below
\begin{equation}
  \label{eq:trav_time}
  \mathbf{d} = \mathbf{r_{XTE}}(t) - \left [ \mathbf{r_{\leftmoon}}(t - \Delta t) - \left ( \mathbf{r_{\oplus}}(t) - \mathbf{r_{\oplus}}(t - \Delta t)\right )\right], 
\end{equation}
where $\mathbf{r_{\oplus}}(t)$ is the Earth position with respect to
the solar system barycenter.
The expression between square bracket represents the Moon position at
time $t - \Delta t$ in the reference frame at time $t$.
Since $\Delta t$ is a function of $d$, Equation(\ref{eq:trav_time})
has to be solved iteratively to find $d$.
We used a fixed point method which converged to the required accuracy
within few steps.
Since all the available lunar surface topographic maps are expressed
with respect to the selenographic reference frame, the next step was
to place the selenographic coordinate system in ICRF/J2000.0 reference
frame.
We were then able to project the lunar surface topography as seen by
\rxte on the plane of sky at both eclipse ingress and egress epoch (see
Figure \ref{fig:ecl_inset}).
The coordinates of the intersection point (inset in Figure
\ref{fig:ecl_inset}) are reported in Table \ref{table1}.

\section{Observations and data analysis} 
The source was discovered on 2010 October 10 \citep{Bordas_atel_10},
and \rxte started the observation of the source on October 13.
The lunar eclipse is present in the first pointed observation of
\igrjc (ObsId 95437-01-01-00) performed by the Proportional Counter
Array (PCA)/\rxte instrument.
The data are available in good-xenon packing mode, with maximum
temporal ($1 \mu$s) and energy resolution (256 channels).
For the kind of analysis we performed, photon arrival times in data
where not reported to the solar system barycenter, since we are
interested in the instant at which photons reach the spacecraft.
Fine clock corrections were not applied since they have negligible
effect in the present analysis.
Indeed, their magnitude ($< 60\mu$s) is two orders of magnitude
smaller than the uncertainties on ingress and egress epochs we
obtained.

In the following, a detailed discussion of all the uncertainties is
given.

\subsection{The Moon} \label{sec:moon}
In the determination of the source position, the sources of error
related to the Moon are two: the positional error of the Moon
barycenter and the uncertainty on the knowledge of the Lunar
topography.

\subsubsection{Uncertainty on the Moon Position} 
In our work, we adopted the Moon position with respect to the Earth as
reported in DE421/LE421 ephemeris \citep{Folkner_08,Williams_08}.

The uncertainty at $1\sigma$ confidence level (hereafter c.l.) in the
lunar orbit for DE421 (taking into account possible systematic errors,
not just formal uncertainties from the least-squares fit) is about 5 m
in right ascension, 2.4 m in declination, and 0.5 m in range
(W. F. Folkner 2012, private communication).

The position of Earth and Moon barycenters, and the spatial
orientation of their spin axis as a function of time was obtained
using the JPL's NAIF SPICE Toolkit.\footnote{Version N0064
  \url{http://naif.jpl.nasa.gov/naif/toolkit.html}.}
We sampled these positions every 15 s.

During the iterative process described in Section \ref{sec:technique},
we interpolated between these positions with a cubic spline.

\subsubsection{Uncertainty on the Moon Topography}
The topography of the lunar surface was measured by Earth-observation
and lunar missions.
The most accurate topographic maps of the moon was obtained by laser
ranging altimetry made by probes orbiting the Moon.

The very first one of such a topographic lunar map was done by the
NASA Clementine mission \citep{Nozette_94}, with a radial precision of
$\sim 40$ m.
In this work, we analyzed the topographic data from the laser
altimetry instrument LALT on board the Japanese lunar explorer Kaguya
(SELENE) \citep{Araki_09, Araki_10}, in particular the global grid
topographic data of the Moon (LALT\_GGD\_NUM, ver. 1, hereafter LALT
data).
The LALT data are referenced to the sphere of 1737.4 km radius based
on the gravity center of the Mean Earth/Polar Axis body-fixed
coordinates of the Moon.\footnote{For details on the lunar coordinate
  system adopted, see
  \url{http://lunar.gsfc.nasa.gov/library/LunCoordWhitePaper-10-08.pdf}.}
The LALT data grid resolution is 0.0625 (1/16) deg \citep{Araki_09,
  Araki_10, Fok_11}. The radial topographic error is 4.1 m
\citep[$1\sigma$ c.l.,][]{Araki_09}.

\subsection{Uncertainties on Ingress and Egress Epochs}
\begin{figure} 
  \resizebox{\hsize}{!}{\includegraphics{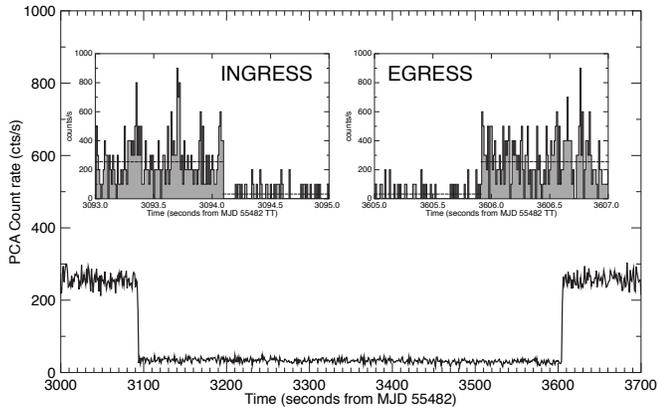}}
  \caption{Light curve of PCA data around the eclipse of \igrjc. The
    light curve is binned at 1 s. In the two insets the light curve of
    eclipse ingress (left) and egress (right) binned at 10 ms are
    reported.\label{fig:eclipse}}
\end{figure}

During the observation in which the lunar occultation occurred, the
source flux was nearly constant with the exception of a type I burst
(outside the time interval shown in Figure \ref{fig:eclipse}).
Excluding the type-I burst, the average count rates outside and during
the eclipse where 254.3(9) and 36.2(9) counts s$^{-1}$, respectively
(number in parenthesis indicates 1$\sigma$ errors on the last digit).
These values are obtained fitting the light curve (binned at 1 s)
inside and outside the eclipse with a constant models.

To obtain an accurate determination of the epoch of eclipse ingress
and egress, we binned the light curve at 100$\mu$s (on such short time
intervals no more than one event occurred in each bin, implying that
the events can be considered unbinned, in practice) and fitted
separately eclipse ingress and egress with step (Heaviside) functions.

Due to the photon paucity, a least-square method was not suitable to
fit the ingress and egress epochs (see insets in Figure
\ref{fig:eclipse}).
We therefore used an unbinned maximum likelihood method, which is more
suitable in such cases \citep{Bevington_03}.
We have taken advantage from prior knowledge of the count rate inside
and outside eclipse, so that the fit is mono-parametric with the only
parameter the transition time.
We fitted separately ingress and egress, considering for each fit the
data in a 10 s time interval centered on ingress (or egress).
To estimate the uncertainty to be associated to the method, we
simulated 10,000 data set with the same characteristics of the
original one, and studied the distribution of the best-fit values.
The confidence interval obtained is asymmetric, reflecting the large
difference between statistics inside and outside eclipse.
The $1\sigma$ uncertainty results to be 7 ms on the high count rate
side, while is 9 ms on the low count rate side.
The ingress and egress epochs and the associated $1\sigma$ c.l. are
reported in Table \ref{table1}.
\begin{table}
  \begin{center}
    \caption{Position of \igrjc and Times of Eclipse
      Ingress/Egress.\label{table1}}
    \begin{tabular}{lc} 
      \tableline 
      \tableline 
      Parameter & Value \\ 
      \tableline 
      \\
      R.A. (ICRF/J2000.0) & \(17^{\rm h}\, 48^{\rm m}\, 04\fs8245(26) \) \\ 
      \\
      Decl. (ICRF/J2000.0) & \(-24\degr\, 46'\, 48\farcs88(4)\) \\ 
      \\
      \tableline 
      \\
      $T_{\mathrm in}$ (MJD TT) & 55482.03581125$^{+10}_{-8}$ \\ 
      \\
      $T_{\mathrm eg}$ (MJD TT) & 55482.04173510$^{+8}_{-10}$ \\ 
      \\
      \tableline
    \end{tabular}
    \tablecomments{Numbers in parenthesis are 1$\sigma$ c.l. errors on
      the last digit(s).}
  \end{center}
\end{table}

\subsection{Uncertainties on Spacecraft Position}
A complex part of the analysis was the determination of the
\textit{Rossi} XTE spacecraft positional errors.
In the following, a brief description of how the spacecraft ephemeris
are derived is given.

The \textit{Rossi} XTE spacecraft positions (and velocities) are
stored in the orbit files as Cartesian coordinates with a sampling
rate of 60 s.
As reported in \cite{Jahoda_06}, the position is obtained from the
spacecraft ephemeris, estimated by the Goddard Flight Dynamics
Facility by fitting the spacecraft position on a 2/3 days basis.
This procedures implies that orbit ephemeris of contiguous orbitfiles
are more or less correlated.
In each orbit file, the spacecraft best-fit ephemeris is also
extrapolated forward by 10 hr, overlapping with the next day solution.
The overlap between two contiguous orbit files is not perfect,
reflecting the orbit variations due to the solar radiation pressure,
and the statistical uncertainties in the determination of the
spacecraft ephemeris.
To estimate the statistical uncertainty on the spacecraft position,
following \cite{Jahoda_06}, we analyzed the distribution of the
distance between the spacecraft positions reported in the overlapping
time intervals from the orbit number 5000 (54354 MJD) to orbit number
6203 (55556.0 MJD), covering a time span of 3.2 yr around the eclipse
epoch.
This distribution shows three peaks, centered on 0, $\simeq 0.01$ m,
and $\simeq 50$ m, respectively, reflecting the aforementioned degree
of correlation between contiguous orbit files.

In line with what is discussed in \cite{Jahoda_06} and following a
discussion with C. B. Markwardt (2012, private communication), as an
upper limit on the uncertainty on the spacecraft position, we
considered the statistical distribution of the $\simeq 50$ m peak.

From this distribution we derive the following confidence intervals
for the \textit{Rossi} XTE position in the considered time interval:
48 m at $1 \sigma$, 121 m at $2 \sigma$, 311 m at $99 \%$, and 970 m
at $3 \sigma$.

\section{Discussion and conclusions}
As described in Section \ref{sec:technique}, we were able to determine
the position of the source as intersection of the lunar rim as seen by
\rxte at eclipse ingress and egress (see Figure \ref{fig:ecl_inset}).
In the following, to calculate the overall uncertainty of \igrjc
position, we consider all the source of error as independent.
The sources of uncertainty can be subdivided in two kind: temporal
(ingress and egress epoch, fine clock corrections) and spatial
(spacecraft and lunar position, lunar topography).

The position on the plane of sky of the source is given by the
direction of the segment joining the spacecraft and the point on the
lunar surface in which the eclipse ingress and egress of the X-ray
source occurred.
This implies that the angular uncertainty on the direction of such a
segment is obtained from the ratio of the overall positional
uncertainty on a plane perpendicular to the source direction and the
Moon--spacecraft distance.
The overall uncertainty on the plane of sky is obtained by summing in
quadrature all the spatial uncertainties involved.

As reported in Section \ref{sec:moon}, the positional uncertainty on
the position of the Moon barycenter are 5 m in R.A., and 2.4 m in
decl.
The corresponding angular uncertainty, calculated considering an
average Moon--spacecraft distance of \np{3.844}{5} km,\footnote{The
  Moon--spacecraft distance is 383418.495 km at ingress, and
  385476.095 km at egress. In our calculations, we considered the
  average.} is of $2.7$ milliarcseconds (hereafter mas) in R.A. and
$1.3$ mas in declination.
The error associated to the lunar topography adopted is the radial
uncertainty of 5 m.
The angular uncertainty we obtain is $2.7$ mas in both axes.

Using a Monte Carlo technique, we estimated a confidence interval of
$^{+9}_{-7}$ ms and $^{+7}_{-9}$ ms for ingress and egress,
respectively.
To convert these temporal uncertainties in spatial uncertainties on a
plane perpendicular to the source direction, we multiply them by
Moon--spacecraft relative speed projected on the plane of sky
$\mathbf{v_{\mathrm rel}}$.
The moduli of these velocity vectors are: $v^{\mathrm R.A.}_{\mathrm in} = 5.70$ km
s$^{-1}$, $v^{\mathrm Decl.}_{\mathrm in} = 2.26$ km s$^{-1}$, $v^{\mathrm R.A.}_{\mathrm eg} = 2.72$ km
s$^{-1}$, $v^{\mathrm Decl.}_{\mathrm eg} = 4.71$ km s$^{-1}$, where the subscripts $\mathrm in$
and $\mathrm eg$ indicate ingress or egress epoch, respectively, while the
superscripts RA and Dec indicate the corresponding axis in the
plane of sky.
Proceeding as described above, we obtain for R.A. and decl. the
following $1 \sigma$ confidence intervals: $\Delta\alpha_{\mathrm in}
= {\phantom{.}^{+27}_{-21}}$ mas, $\Delta\alpha_{\mathrm eg} =
{\phantom{.}_{-13}^{+10}}$ mas, $\Delta\beta_{\mathrm in} =
{\phantom{.}_{-11}^{+8}}$ mas, $\Delta\beta_{\mathrm eg} =
{\phantom{.}_{-18}^{+23}}$ mas.
The uncertainties in ingress and egress where summed in quadrature.

Finally we considered the uncertainty in the spacecraft position,
which is 48 m at $1 \sigma$ confidence level, corresponding to an
uncertainty of $26$ mas on both axes.\footnote{We note that, formally,
  we should project the spacecraft position uncertainty on the plane
  of sky.
  Such a procedure implies that the uncertainty to be associated
  should be a $\sim$$\sqrt{2}$ factor smaller.
  However, to be conservative, we prefer to use the original
  uncertainty.}

To summarize, we report in Table \ref{table2} all the uncertainties
discussed above at $1\sigma$ c.l. on R.A. and decl., together with the
grand total, obtained summing them in quadrature.
\begin{table}
  \begin{center}
    \caption{Error Budget\label{table2}}
    \begin{tabular}{lcc} 
      \tableline 
      \tableline 
      Parameter & $\Delta$ R.A. (mas)&  $\Delta$ Decl. (mas) \\ 
      \tableline 
      Moon ephemeris & $\pm 2.7$ & $\pm 1.3$ \\
      \textit{Rossi} XTE ephemeris & $\pm 26$ & $\pm 26$ \\
      Eclipse ingress and egress epochs & $_{-25}^{+29}$ & $_{-21}^{+24}$ \\
      Moon topography radial precision & $\pm 2.7$ & $\pm 2.7$ \\
      \tableline
      Grand total & $_{-36}^{+39}$ & $_{-33}^{+35}$\\
      \tableline
    \end{tabular}
    \tablecomments{Statistical errors are intended to be at $1\sigma$
      confidence level.}
  \end{center}
\end{table}

Our result confirms the association made by \cite{Pooley_atel_10} of a
quiescent X-ray source observed by \cite{Heinke_06} in the globular
cluster Terzan 5.
Indeed our position is $0\farcs1$ from the quiescent X-ray source
detected by \cite{Heinke_06} in a 40 ks \chandra X-ray observation of
the globular cluster Terzan 5.
Note that the accuracy on the source position reported by
\cite{Heinke_06} is $0\farcs06$ ($1\sigma$ c.l.), slightly worse than
the uncertainty derived form the Moon occultation ($\simeq
0\farcs04$).
\begin{figure*} 
  \plotone{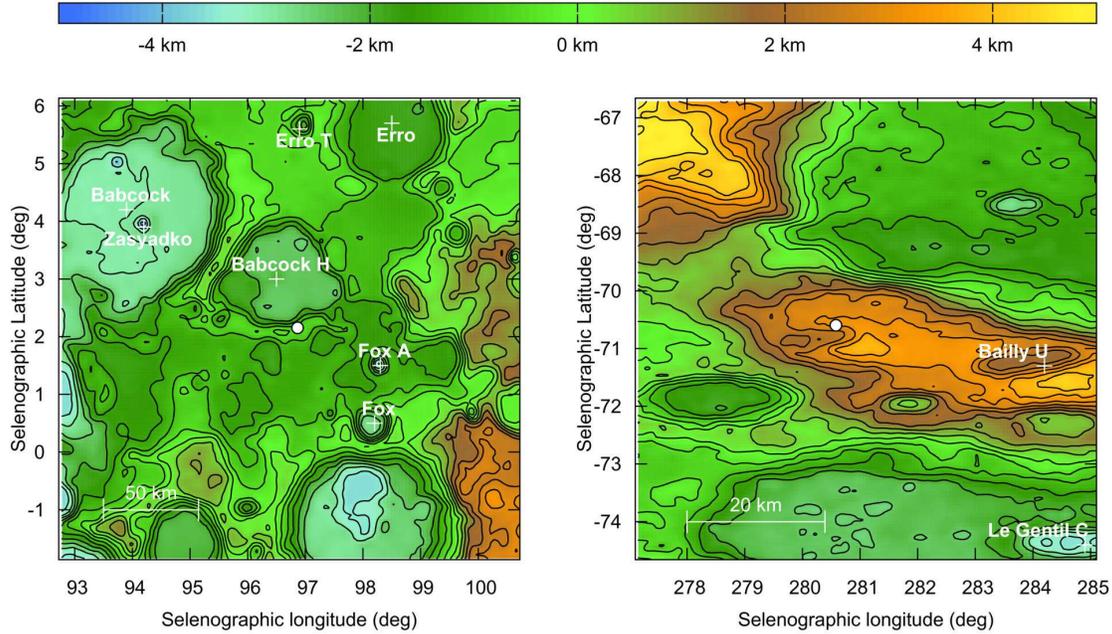}
  \caption{Moon topography of the $8^\circ \times 8^\circ$ regions
    around the eclipse ingress (left panel) and egress (right panel)
    points. Each map is centered on the direction of the moon radius
    orthogonal to the source direction. The white circles identify the
    points on the lunar surface where the ingress and egress took
    place. The contour interval is 500 m. The white crosses indicated
    the most important lunar craters in the
    regions.\label{fig:kaguya_maps}}
\end{figure*}
{Moreover, V. Testa et al. (2012, in preparation) use the high
  precision on the X-ray source position obtained with Moon
  occultation to identify a possible near-IR counterpart even in a
  extreme crowded field of view as the center of a globular cluster.}

{In the following we discuss the applicability of this method to
  present and future X-ray missions. In particular we briefly compare
  \rxte with \chandra, \xmm, \nustar, and LOFT discussing the
  positional accuracy obtainable with this technique.}
{As discussed in Section \ref{sec:technique}, the accuracy on the
  source position depends on the accuracy in the ingress and egress
  epochs, which in turn depends on the source and background count
  rates through the following relation $\delta t \simeq (1 + R_{\rm
    bkg}/R_{\rm src}) / R_{\rm src}$ (at $1\sigma$ confidence level),
  $R_{\rm src}$ and $R_{\rm bkg}$ are the count rate of the source and
  the background, respectively. Therefore the total uncertainty, at
  $1\sigma$ level, in source position, $\sigma_{\theta}$, can be
  written as}
\begin{equation}
  \label{eq:accuracy}
  \sigma_{\theta}^2  \simeq \frac{\mathbf{v_{\rm rel}^2}}{d^2} {\left (\frac{1 + {R_{\rm bkg}}/{R_{\rm src}}}{R_{\rm src}}\right)}^{\!\!2} + \frac{\sigma^2_{\leftmoon} + \sigma^2_{\rm spc}}{d^2},
\end{equation}
{where $\sigma_{\leftmoon}$ and $\sigma_{\rm spc}$ are the
  uncertainties in the Moon rim and spacecraft position, $d$ is the
  Moon--spacecraft distance.}
{Since solar system ephemeris, and in particular Lunar ephemeris, are
  already at an accuracy of 5 m, while present lunar mission LRO/LOLA
  \citep{Smith_10} is mapping its surface with an overall accuracy
  $\lesssim 1$ m, the precision is mainly limited by the uncertainties
  on spacecraft position and on the ingress and egress epochs.}
The first one can easily be improved in present and future spacecrafts
equipped with GPS.
{This means that the leading term in the uncertainty
  $\sigma_{\theta}$ is the uncertainty on the ingress and egress
  epochs.}
{For source dominated targets ($R_{\rm bkg} / R_{\rm src} << 1$), the
  positional accuracy obtainable with this technique scales as the
  inverse of the source count rate, which. for a given spectral shape,
  depends on the effective area and energy band of the X-ray
  satellite.}
{The effective area at 5 keV for present X-ray satellite are 1000
  cm$^2$ for each Proportional Counter Unit in \rxte \citep{Bradt_93},
  800 cm$^2$ for EPIC/pn in \xmm \citep{Jansen_01}, 300 cm$^2$ for
  \chandra/EPIC \citep{Weisskopf_02}, while the effective are of
  \nustar will be 700 cm$^2$ \citep{Harrison_10} and of LOFT/LAD will
  be 110,000 cm$^2$ \citep{Feroci_11}.}
{Unfortunately, such a technique cannot be used with \xmm, since it
  has a constraint on the angle between the pointing direction and the
  Moon ($> 22^\circ$), because of the limitations imposed by the
  Optical Monitor.}
{For bright sources, continuous clocking mode must be used with
  \chandra, in order to avoid pile-up photon losses.}
A future X-ray mission with a $\sim$10 m$^2$ effective area such as
LOFT \citep{Feroci_11} will then permit to go at a level below 5 mas.

\textit{Note added in proof} We have re-calculated the position of the
11-Hz pulsar IGR J17480-2446 in Terzan 5 using data from LRO/LOLA for
the Lunar topography. These data have an intrinsic uncertainty of 1 m
on the Lunar surface. In this case we find the following source
position: $17^{\rm h}\, 48^{\rm m}\, 04\fs8216(26),-24\degr\, 46'\,
48\farcs88(4)$.  This position differs from the one obtained from
Kaguya/LALT by 43 mas, which is about 1 sigma the error box.

\begin{acknowledgements} 
  This work is supported by the Italian Space Agency, ASI--INAF
  I/088/06/0 and I/009/10/0 contracts for High Energy Astrophysics.
  This research made use of the lunar orbiter SELENE (KAGUYA) data of
  JAXA/SELENE.
  We thank W. Folkner for having provided useful information on JPL
  solar system ephemeris, C. B. Markwardt for having provided useful
  information on \textit{Rossi} XTE orbit ephemeris, B. Saitta and
  A. De Falco for useful discussions on statistical methods.
  A. Papitto acknowledges the support by the grants AYA2009-07391 and
  SGR2009-811, as well as the Formosa program TW2010005 and iLINK
  program 2011-0303.
  E. Egron acknowledges the support of the Initial Training Network
  ITN 215212: Black Hole Universe funded by the European Community.
\end{acknowledgements}

\bibliography{ms}

\end{document}